\documentclass[aps,prl,groupedaddress]{revtex4-1}
\usepackage{tikz,amssymb}
\newcommand{\id}{\mbox{\rm1\hspace{-.2ex}\rule{.1ex}{1.44ex}}
   \hspace{-.82ex}\rule[-.01ex]{1.07ex}{.1ex}\hspace{.2ex}}

\begin{document}


\title{The ground state energy of the mean field spin glass model
}


\author{F.\ Koukiou}
\email[]{flora.koukiou@u-cergy.fr}
\affiliation{Laboratoire de physique th\'eorique et mod\'elisation
\footnote{UMR 8089, CNRS}
\\
Universit\'e de Cergy-Pontoise
F-95302 Cergy-Pontoise}


\date{\today}

\begin{abstract}

We establish a  functional equation relating  the Gibbs measure at a particular low temperature with the one at temperature 1. This equation enables us to calculate  the limiting  free energy
 of the Sherrington-Kirkpatrick spin glass model at this particular value of low temperature without making use of the replica method.  We get additionally a sharper
 lower  bound for the ground state energy 
   $\epsilon_0$  
\[\epsilon_0\geq 
 -0.7833\cdots,\]
 close to the replica symmetry breaking and numerical simulations  values.
\end{abstract}


\maketitle



\section{Introduction and main result}

During the last decade, mean field models of spin glasses
have motivated  increansingly many
studies by      physicists and mathematicians \cite{ALR, G, GT, K1,
 P, PS, T1}.
The rigorous understanding 
 of the infinite volume limit of thermodynamic quantities 
remained quite insufficient until the recent  breakthrough obtained 
by  Guerra and Toninelli \cite{GT} 
on their  existence and uniqueness.  This major discovery 
followed by  several  important  results \cite{ASS, G}
 providing
a mathematical interpretation of the original formulae proposed 
by Parisi \cite{P} on the basis of heuristic arguments.

In this note, without making use of the replica approach, we calculate,  for a particular value of the (low) temperature,  the  limiting free energy  of the
 Sherrington-Kirkpatrick model
and obtain  a  lower bound  for the density of the  ground state energy.  Although the limiting free energy is given, for the whole low temperature region, by the  rather complicated Parisi formula,  
we obtain,  for a given  temperature, a very  simple   expression.   This allows  an improvement of all known rigorous bounds for the ground state energy.

We first recall some basic definitions.
Suppose that a finite set of $n$ sites is given.
Let $\sigma_i\in \{1, -1\}$ be the spin variable on the site $i$ and $\sigma$
a 
generic configuration in the configuration space $\Sigma_n=\{-1,1\}^n$. 
The finite volume Hamiltonian of the model 
is given by the following real-valued function on $\Sigma_n$
\[H_n(\sigma)=-\frac{1}{\sqrt{n}}
\sum_{1\leq i<j\leq n}J_{ij}\sigma_i\sigma_j,\]
where the family of couplings $J_{ij}$ are 
independed centered Gaussian random variables of variance $1$. 
Note however, that the function $H_n$ can equivalently be defined directly on the infinite configuration space $\Sigma_\infty=\{-1,1\}^\mathbb{N}$ and the infinite collection of random variables $J=(J_{ij})_{i,j\in\mathbb{N}}$ just by the  trivial modification in the definition of $H_n$:
\[H_n(\sigma)=-\frac{1}{\sqrt{n}}
\sum_{1\leq i<j\leq n}J_{ij}\sigma_i\sigma_j\id_{\Sigma_n}(\sigma).\]
This remark will be instrumental in the course of the proof since only changes the sequence of functions $(H_n)_{n\in\mathbb{N}}$ and not the configuration and environment spaces.
Henceforth, the Hamiltonian will be defined on these infinite spaces.

For the  inverse temperature $\beta=\frac{1}{T}>0$, 
the disorder dependent  partition function $Z_n(\beta)$, 
  is defined   by 
  \[Z_n(\beta, J)=\sum_{\sigma}\exp (-\beta H_n(\sigma,J)).\]
Moreover,  if
$E_J$ denotes the expectation with respect to the randomness $J_{ij}$,
 it is very simple to show that  $E_JZ_n(\beta,J)=2^n e^{\frac{\beta^2}{4}(n-1)}$.  
 
We denote by $\mu_{n,\beta}(\sigma|J)$, the  corresponding  Gibbs
probability measure, conditionned 
on fixed randomness: 
\[\mu_{n,\beta}(\sigma|J)=
e^{-\beta H_n(\sigma,J)}/Z_n(\beta, J),\]
and, by  $S(\mu_{n,\beta}(\sigma|J))$,  its entropy,  defined by
$S(\mu_{n,\beta}(\sigma|J))=-\sum_\sigma\mu_{n,\beta}(\sigma|J)
 \log  \mu_{n,\beta}(\sigma|J)$.
 
For fixed randomness, the  real functions  
\[f_n(\beta)=\frac{1}{n}E_J\log Z_n(\beta,J)\]and
\[\bar{f}_n(\beta)=\frac{1}{n}\log E_J Z_n(\beta,J),\]
define  the quenched average of the free energy per site and 
 the annealed  specific free energy respectively.

 The ground state energy density $-\epsilon_n(J)$ is defined by
\[-\epsilon_n(J)=\frac{1}{n}\inf_{\sigma\in\Sigma_n} H_n(\sigma,J).\]

For the low temperature region ($\beta>1$),
 the $J$-almost sure  existence 
 of the infinite volume limits
  \[\lim_{n\rightarrow\infty}f_n(\beta)=f_\infty(\beta),\] and, 
\[-\lim_{n\rightarrow\infty}\epsilon_n(J)=
\lim_{\beta\rightarrow\infty}\frac{f_\infty(\beta)}{\beta}=-\epsilon_0\]
was first  proved by Guerra and Toninelli \cite{GT}. More recently, 
 Aizenman, Sims and Starr \cite{ASS} gave a clear mathematical interpretation
 of the limit  $f_\infty(\beta)$ in terms of the variational
formula proposed by Parisi.

In the following section we prove  the \\ 

\noindent
{\bf Theorem~:}
{\it Let  $\beta_*=4 \log 2=2.77258\cdots$  Almost surely, the infinite volume limit $f_\infty(\beta_*)$ is given by 
\[f_\infty(\beta_*)=\lim_{n\rightarrow\infty}\frac{1}{n}E_J\log Z_n(\beta_*,J)
=\frac{\beta_*^2}{4}+\frac{1}{4}.\]}

A directly  related result is the following corollary, which improves all the rigorous 
lower bounds for the ground state energy.\\

\noindent
{\bf Corollary~:}
{\it Almost surely,  the ground state energy 
 density  of the Sherrington-Kirkpatrick  
spin glass model is bounded  by}
\[\epsilon_0\geq  = -0.7833\cdots.\]

\section{\bf Proof of the main result}

Notice first  that for all  $\beta>0$,  the limit   $f_\infty(\beta)$
exists and it is a convex function of $\beta$ \cite{GT}.
Let $\beta=\beta_1\equiv 1$. From the high temperature results \cite{ALR}, we 
have,   almost surely, that 
\begin{equation}\label{eq:1}
f_\infty(\beta_1)=\lim_{n\rightarrow\infty}\frac{1}{n}
E_J\log Z_n(\beta_1,J)=
\bar{f}_\infty(\beta_1)=\lim_{n\rightarrow\infty}\frac{1}{n}\log E_JZ_n(\beta_1,J)=
\log 2 +\frac{\beta_1^2}{4}=\log 2+\frac{1}{4}.
\end{equation}
The following figure \ref{fig:beta}  illustrates the definition of the inverse temperature $\beta_*$;  the annealed free energy $\bar{f}_\infty(\beta)=
\log 2+\frac{\beta^2}{4}$ 
is plotted as a function of $\beta$ and the straight line is defined  by $\frac{\beta}{\beta_1} f_\infty(\beta_1)=\beta  f_\infty(\beta_1)$.

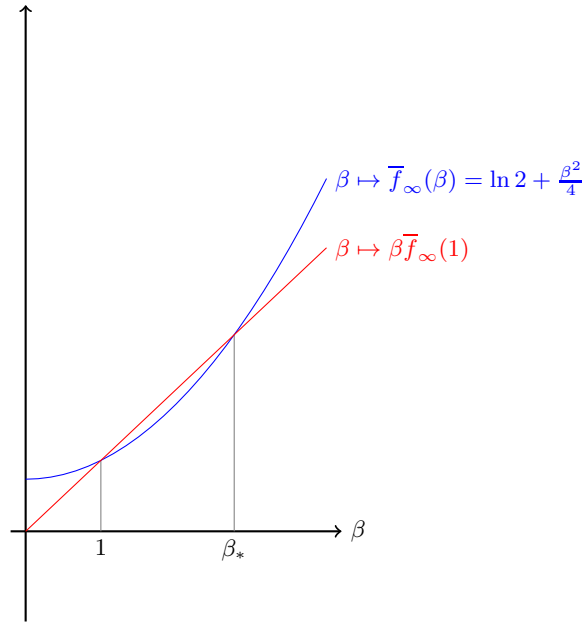
\begin{figure}[h]
\begin{center}
\begin{tikzpicture}[domain=0:4]
 
    \draw[->, thick,color=black] (-0.2,0) -- (4.2,0) node[right] {$\beta$};
    \draw[->, thick, color=black] (0,-1.2) -- (0,7);
   \draw[-, color=gray]    (1,0) -- (1, 0.943147) ;
   \draw[-] (1,0)--(1,0.001) node[below] {$1$};
   \draw[-] (2.772588,0)--(2.772588,0.001) node[below] {$\beta_*$};
    \draw[-, color=gray]    (2.772588,0) -- (2.772588,2.614959 ) ;
    \draw[color=blue]  plot[id=annealed] function{log(2)+0.25*(x**2)}
        node[right] {$\beta\mapsto\overline{f}_\infty(\beta) =\ln 2+ \frac{\beta^2}{4}$};
    \draw[color=red] plot[id=linear] function{x*(log(2)+0.25)} 
        node[right] {$\beta \mapsto \beta\overline{f}_\infty(1)$};
  \end{tikzpicture}
\end{center}
\caption{\label{fig:beta} 
The value  $\beta_*=2,77258\cdots$,  is given by the intersection of  the graph of the 
  annealed free energy $\bar{f}_\infty(\beta)$ 
  with the straight line  $\beta f_\infty(1)$.}
\end{figure}

One can remark  that, for   $\beta_*=2,77258\cdots=4\log 2$,  the annealed free energy $\bar{f}_\infty(\beta_*)$ is simply related to $f_\infty(\beta_1)$  via  the following relation
\begin{equation}\label{eq:2}
\bar{f}_\infty(\beta_*)=\frac{\beta_*^2}{4}+\log 2=\frac{\beta_*}{\beta_1}(\frac{\beta_* \beta_1}{4}+
\frac{\beta_1}{\beta_*}\log 2)=\frac{\beta_*}{\beta_1}
(\log 2+\frac{1}{4})=\frac{\beta_*}{\beta_1} f_\infty(\beta_1).
\end{equation}

By making use of  this remark we  define  the 
 Gibbs probability  measure  
$\mu_{n,\beta_*}(\sigma|J)$  by the functional equation  
\begin{equation}\label{eq:3}
\mu_{n,\beta_*}(\sigma|J):=\frac{\exp(\beta_* H_n(\sigma, J))}{Z_n(\beta_*,J)}=\mu_{n,\beta_1}^{\frac{\beta_*}{\beta_1}}(\sigma|J)
\frac{Z_n^{\frac{\beta_*}{\beta_1}}(\beta_1,J)}{Z_n(\beta_*,J)},
 \end{equation}
induced by the mapping $T:\exp(\beta_1 H_n(\sigma, J))\mapsto (\exp(\beta_1H_n(\sigma, J)))^{\beta_*/\beta_1}$ among Boltzmann factors.
Since $\mu_{n,\beta_*}$ is a probability on the configuration space, summing over $\sigma$, we have  indeed    
\begin{equation}\label{eq:4}
\lim_{n\rightarrow \infty}\frac{1}{n}E_J
 \log \sum_\sigma\mu_{n,\beta_1}^{\frac{\beta_*}{\beta_1}}(\sigma|J)
=-\alpha_\infty(\frac{\beta_*}{\beta_1}),
\end{equation}
where  
 the limit $\alpha_\infty(\frac{\beta_*}{\beta_1})$
gives the deviation of the  free energy $f_\infty(\beta_*)$ from its mean value:
\begin{eqnarray*}\label{eq:5}
\alpha_\infty:=\alpha_\infty(\frac{\beta_*}{\beta_1})
& = &
\lim_{n\rightarrow \infty}\frac{\beta_*}{\beta_1}\frac{1}{n}E_J\log Z_n(\beta_1,J)-
\lim_{n\rightarrow \infty}\frac{1}{n}E_J\log Z_n(\beta_*,J)\\
& = &
\frac{\beta_*}{\beta_1}f_\infty(\beta_1)-f_\infty(\beta_*)\\
& = &\bar{f}_\infty(\beta_*)-f_\infty(\beta_*).
\end{eqnarray*}
The reader can remark that  the value $\beta_1$ fixes the temperature scale; the temperature $\beta_*$ is expressed in units where $\beta_1\equiv 1$. Therefore,  $\frac{\beta_*}{\beta_1}$ is a non dimensional quantity.

 We can now  calculate the limit $f_\infty(\beta_*)$. The idea is the following.
 For $\eta>0$, 
 consider the set $S$ of configurations $\sigma$ where the 
 Boltzmann factor is close to  its mean value:
 $|\exp(-\beta H_n(\sigma,J))- \exp(\frac{\beta^2}{4}(n-1)|\leq \eta.$
 For $\beta=1$, the  set  $S$ is of full measure and moreover, 
the Gibbs measure behaves, at the thermodynamic limit, as
$\lim_{n\rightarrow\infty}\frac{1}{n} \log \mu_{n,\beta_1}(\sigma|J)=
\frac{\beta^2_1}{4}-f_\infty(\beta_1)=-\log 2$.

 Taking now the image under $T$, we have the following behaviour of the  limit
\begin{eqnarray*}\label{eq:6}
\lim_{n\rightarrow\infty}\frac{1}{n} \log \mu_{n,\beta_*}(\sigma|J) & = &
\frac{\beta_* \beta_1}{4}-f_\infty(\beta_*)\\
& = &
-\frac{\beta^2_*}{4}+\alpha_\infty,
\end{eqnarray*} 
{\it i.e.}  this image differs from the limit
$-f_\infty(\beta_*)$ by $\frac{\beta_*\beta_1}{4}=\log 2$.
Now, from the previous equation and  the  easily verified  relation
\begin{equation}\label{eq:7}
\frac{\beta^2_*}{4}-(\alpha_\infty+\frac{\beta_1^2}{4}))=f_\infty(\beta_*)-f_\infty(\beta_1),
\end{equation}
we can deduce the behaviour of the limit $\alpha_\infty$ by making use of the following argument. 
First, the  difference of the limit $-\frac{\beta_*^2}{4} +\alpha_\infty$  from $-\log 2 +\alpha_\infty$ is  given  by  $(\beta_*-1)\log 2.$
Moreover, the difference between the limiting values $-\frac{\beta_*^2}{4} +\alpha_\infty+\frac{\beta_1^2}{4}$ and 
$-\log 2 +\alpha_\infty+\frac{\beta_1^2}{4}$ is also $(\beta_*-1)\log 2.$

\begin{figure}[h]
\begin{center}
\begin{tikzpicture}[scale=5]
    \draw[<-, thick,color=black] (-1.2,0) -- (0,0);
 \draw[color=black]   (-1.5,0)  node [above] {\scriptsize{$\lim_{n\rightarrow\infty} \frac{1}{n}\log\mu_{n,\beta_1} (\sigma)$}};
    \draw[->, thick, color=black] (0,0) -- (0,-2.3);
    \draw[color=blue] (-0.99,-2.3017156571)--(-0.159831559,0);
    \draw[color=red] (-0.693314718,0) -- (-0.693314718,-1.4786648777) -- (0,-1.4786648777); 
     \draw[color=red] (-0.9431471806,0) -- (-0.9431471806,-2.1718120583) -- (0,-2.1718120583); 
      \draw[color=black] (-0.693314718,0)  circle(0.01pt) node[above] {\scriptsize{$-\log2$}}; 
       \draw[color=black] (-0.94314718,0)  circle(0.01pt) node[above] {\scriptsize{$-f_\infty(\beta_1)$}}; 
\draw[color=black] (0,-1.4786648777)  circle(0.01pt) node[right] {\scriptsize{$-\beta_*^2/4+\alpha_\infty=-f_\infty(\beta_*)+\log 2$}}; 
  \draw[color=black] (0,-2.1718120583) circle(0.01pt)
  node[right] {\scriptsize{$-f_\infty(\beta_*)$}}; 
    \draw[color=black] (0,-1.9218120583)  circle(0.01pt)  node[right] {\scriptsize{$-\beta_*^2/4)$}}; 
     \draw[color=black]   (0,-2.28)  node [right] {\scriptsize{$\lim_{n\rightarrow\infty} \frac{1}{n}\log\mu_{n,\beta_*} (\sigma)$}};
  \end{tikzpicture}
\end{center}
\caption{\label{fig:mapping} 
The straight oblique line represents the effect of the mapping $T$ on measures.}
\end{figure}
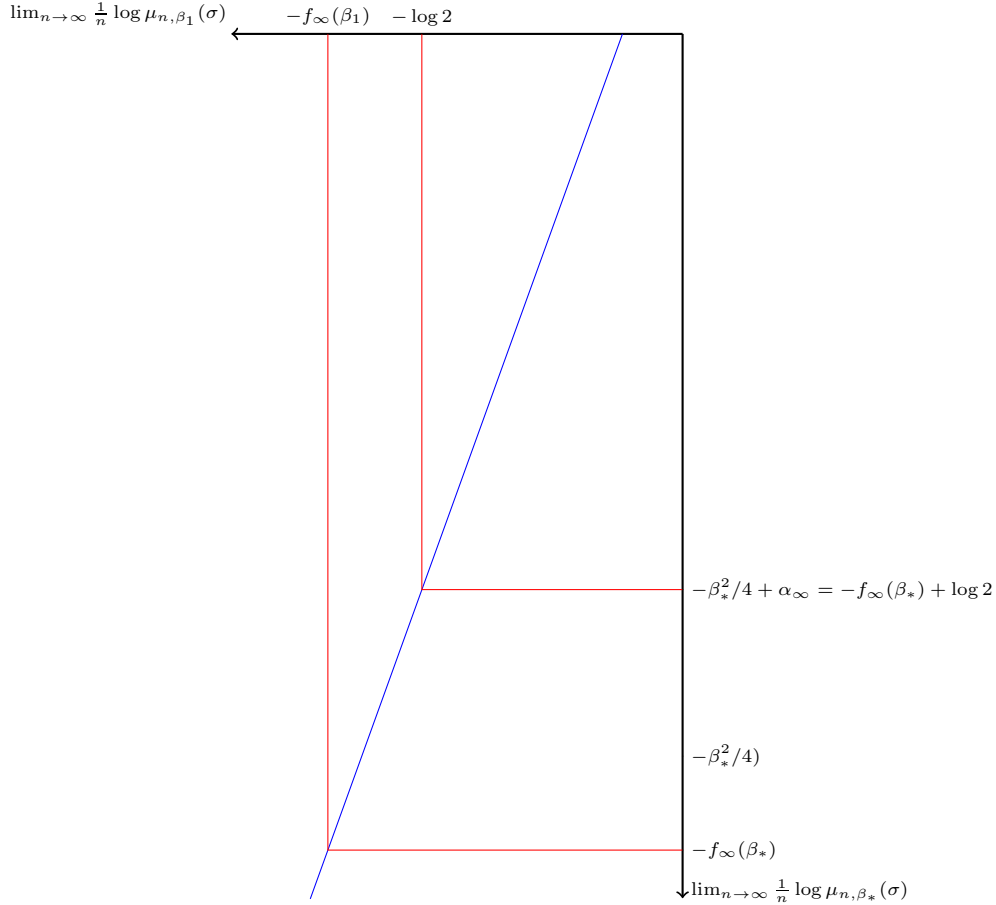

Second, from equation (\ref{eq:7}), since the limit $-\frac{\beta_*^2}{4} +\alpha_\infty+\frac{\beta_1^2}{4}$ is equal to 
$-f_\infty(\beta_*)+f_\infty(\beta_1)$, it follows that 
$-\log 2 +\alpha_\infty+\frac{\beta_1^2}{4}=0$. 
We  have indeed, \begin{equation}\label{eq:10}
\alpha_\infty+\frac{\beta_1^2}{4}
=\log 2= \frac{\beta_1\beta_*}{4},
\end{equation}
and 
\begin{equation}\label{eq:11}
f_\infty(\beta_*)-f_\infty(\beta_1)=(\beta_*-1)\log 2
\end{equation}
as illustrated in figure (\ref{fig:mapping}).

One can now check that the obtained value of the limit 
\begin{equation}
f_\infty(\beta_*)=\beta_* \log 2 +\frac{\beta_1^2}{4}=
\frac{\beta_*^2}{4}+\frac{\beta_1^2}{4}=2.1718\cdots,
\end{equation}
 is   lower than the  spherical model bound   ($2.2058\cdots$).\\

\textbf{Remark 1:}
Our result can also be obtained using a large deviations approach, namely  by comparing the Gibbs measures  at the inverse temperatures  $1$ and $\beta_*$ when the corresponding Boltzmann factors
$\exp(-1H_n(\sigma,J))$ and $\exp(-\beta_*H_n(\sigma,J))$ behave as their mean values 
$\exp(\frac{n-1}{4})$ and $\exp(\frac{\beta_*^2}{4}(n-1))$ respectively.
In this case, the two limits $\lim_{n\rightarrow\infty}\frac{1}{n} \log \mu_{n,\beta_*}(\sigma|J)$  and $\lim_{n\rightarrow\infty}\frac{1}{n} \log \mu_{n,\beta_1}(\sigma|J)$ differ by $\alpha_\infty$.

\textbf{Remark 2:}
The point where these  two limits 
are equal corresponds to the fixed point of the functional equation (\ref{eq:3}) and arises when 
$\lim_{n\rightarrow\infty}\frac{1}{n} \log \mu_{n,\beta_*}(\sigma|J)=-\log 2+\alpha_\infty$ .

We can now obtain a lower bound for the ground state energy density $-\epsilon_n(J)$.
Notice that 
\begin{equation}\label{eq:11}
f_\infty(\beta_*)=-\lim_{n\rightarrow\infty}\frac{\beta_*}{n}\sum_\sigma
H_n(\sigma|J) \mu_{n,\beta_*}(\sigma|J) +s(\mu_{\beta_*})=
\frac{\beta_*^2}{4}+\frac{1}{4},
\end{equation}
where  the  limit
\begin{equation}\label{eq:12}
s(\mu_{\beta_*})=\lim_{n\rightarrow\infty}
\frac{1}{n}S(\mu_{n,\beta_*}(\sigma|J)),
\end{equation}
gives  the (mean) entropy of the Gibbs measure.

We have indeed, by the positivity of the entropy  $s(\mu_{\beta_*}) \geq 0$,   that
\begin{equation}\label{eq:13}\epsilon_0=-\lim_{\beta\rightarrow\infty}\frac{f_\infty(\beta)}{\beta} \geq -\frac{\beta_*}{4}-\frac{1}{4\beta_*}=-0.7833\cdots; 
\end{equation}
close to  the  value $-0.7633\cdots$ obtained by numerical simulations based on the replica  approach. 

\section{Concluding remarks}

In this note  we  obtained, under the assumption of minimal entropy,  a rigorous lower bound for the ground state energy   density  which improves all the previous estimations. 

A last observation concerns  the 
value of the temperature $\beta_*$: it is obtained from the relation (\ref{eq:2})
between the  free energies $\bar{f}_\infty(\beta_*)$ and $f_\infty(1)$; 
moreover, one can readily  check that $\beta_*$   is given by 
 $\beta_*=\beta^2_c$, where
$\beta_c=2\sqrt{\log 2}$ is the critical temperature of the Random- Energy Model (REM) \cite{Der}. The REM  is  defined by $2^n$ energy 
levels $E_i (i=1,\cdots,n)$, a family of random, independent, identically
distributed random variables;  many results are qualitatively the same 
as those of the SK model.  It would be interesting to clarify   this relationship in order to obtain some information on the   behaviour  and properties of the Gibbs measure at low temperatures. 
Both  $\beta_c$ and $\beta_*$ are to be compared with
the value at $\beta_1\equiv 1$, {\it i.e.} the maximum value of $\beta$ where the free energies of the two models coincide. What  we learn by the comparison of the two models is that the Gibbs measure of the SK 
has seemingly a  richer structure
than for the REM.  As a matter of fact, the entropy of the REM vanishes at
$\beta_c$ while the entropy of the SK  model is still strictly positive at this point. We expect  moreover that
the entropy  of the SK model vanishes at $\beta_*$  but this remains  an open problem that is discussed in a forthcoming paper \cite{Entropy}.


\end{document}